\documentclass[conference]{IEEEtran}
\IEEEoverridecommandlockouts
\usepackage{cite}
\usepackage{url}
\usepackage{booktabs}
\usepackage{amsmath,amssymb,amsfonts}
\usepackage{algorithmic}
\usepackage{graphicx}
\usepackage{lipsum}
\usepackage[T1]{fontenc}
\usepackage{textcomp}
\usepackage{tikz}
\usepackage{atbegshi}
\usepackage{xcolor}
\usepackage{csquotes}
\usepackage{comment}
\usepackage{hyperref}
\hypersetup{
colorlinks=true,
linkcolor=blue,
filecolor=blue,
citecolor = blue,
urlcolor=blue,
}
\def\BibTeX{{\rm B\kern-.05em{\sc i\kern-.025em b}\kern-.08em
    T\kern-.1667em\lower.7ex\hbox{E}\kern-.125emX}}
\begin{document}

\title{Deep Learning Approach for Enhancing Oral Squamous Cell Carcinoma with LIME Explainable AI Technique\\
}

\author{
\IEEEauthorblockN{Samiha Islam}
\IEEEauthorblockA{\textit{Dept. of ECE} \\
\textit{North South University}\\
Dhaka-1229, Bangladesh \\
samiha.islam2@northsouth.edu}
\and
\IEEEauthorblockN{Muhammad Zawad Mahmud}
\IEEEauthorblockA{\textit{Dept. of ECE} \\
\textit{North South University}\\
Dhaka-1229, Bangladesh \\
zawad.mahmud1@northsouth.edu}
\and 
\IEEEauthorblockN{Shahran Rahman Alve}
\IEEEauthorblockA{\textit{Dept. of ECE} \\
\textit{North South University}\\
Dhaka-1229, Bangladesh \\
shahran.alve@northsouth.edu}
\and 
\IEEEauthorblockN{Md. Mejbah Ullah Chowdhury}
\IEEEauthorblockA{\textit{Dept. of ECE} \\
\textit{North South University}\\
Dhaka-1229, Bangladesh \\
mejbah.chowdhury@northsouth.edu}
\and 
\IEEEauthorblockN{Faija Islam Oishe}
\IEEEauthorblockA{\textit{Dept. of ECE} \\
\textit{North South University}\\
Dhaka-1229, Bangladesh \\
faija.oishe@northsouth.edu}
}

\maketitle
\begin{abstract}
The goal of the present study is to analyze an application of deep learning models in order to augment the diagnostic performance of oral squamous cell carcinoma (OSCC) with a longitudinal cohort study using the Histopathological Imaging Database for oral cancer analysis. The dataset consisted of 5192 images (2435 Normal and 2511 OSCC), which were allocated between training, testing, and validation sets with an estimated ratio repartition of about 52\% for the OSCC group, and still, our performance measure was validated on a combination set that contains almost equal number (50:50) of sample in this use case as entire database have been divided into half using stratified splitting technique based again near binary proportion but total distribution was around even. We selected four deep-learning architectures for evaluation in the present study: ResNet101, DenseNet121, VGG16, and EfficientnetB3. EfficientNetB3 was found to be the best, with an accuracy of 98.33\% and F1 score (0.9844), and it took remarkably less computing power in comparison with other models. The subsequent one was DenseNet121, with 90.24\% accuracy and an F1 score of 90.45\%. Moreover, we employed the Local Interpretable Model-agnostic Explanations (LIME) method to clarify why EfficientNetB3 made certain decisions with its predictions — to improve the explainability and trustworthiness of results. This work provides evidence for the possible superior diagnosis in OSCC activated from the EfficientNetB3 model with the explanation of AI techniques such as LIME and paves an important groundwork to build on towards clinical usage.
\end{abstract}


\section{Introduction}
Oral squamous cell carcinoma (OSCC) is a malignant tumor originating from the oral cavity, oropharynx, and hypopharynx. It is the most frequently diagnosed form of oral cancer, being over 90\% of all cases. Oral squamous cell carcinoma (OSCC) is a disease that affects various regions of the mouth, including lips, tongue, floor of the mouth, and hard palate. Importantly, early detection greatly enhances the prognosis and chances of survival. In the first place, early lesions in OSCC sometimes remain asymptomatic, representing a late diagnosis at an advanced stage and, consequently, difficult therapy with a poor prognosis.

OSCC exerts a major burden in Asia, having a high frequency of exposure to risk factors (tobacco use, betel quid chewing, genetic assets). These factors result in an increase in oral cancers experienced by Asian countries than Western populations. For example, countries like Bangladesh, India, Pakistan, and Sri Lanka frequently report a large burden of OSCC, making them priority health problems with which to deal. The disease burden advances not only health outcomes but also imposes high costs of care and loss of productivity, highlighting the requirement for efficient and early-stage recognition programs.

Recent progress with deep learning, a subfield of AI, seems to be promising for object detection and classification in OSCC. Deep learning models, and those that use transfer-earning in particular, take advantage of pre-trained neural networks to provide high accuracy with limited data compared to more traditional machine-learning frameworks. This is particularly advantageous in medical imaging that typically does not have large annotated datasets. The transfer learning models are built on top of VGG/Inception/ResNet, and they can generalize the well-disguised discriminative image features from millions of oral histopathological slides that a human expert's naked eye may not be able to detect, hence empowering clinicians with an invincible weapon for early treatment plan intervention in OSCC cases. The early detection landscape of oral cancers may be revolutionized with the integration and appropriate application of AI into diagnostic pathways, leading to possible advances in better management, improved survival rates, etc. The main contributions of this study are:

\begin{itemize}
     \item Our main contribution is finetuning a pre-trained EffiecientNetB3 model for oral squamous cell carcinoma classification. This model outplayed all existing accuracy models for this data set to this point.
     \item The second contribution is to apply LIME XAI to identify the working of the EfficientNetB3 model.
\end{itemize}

\section{Related Work}
In the last few years, a slew of research has been carried out with deep learning (DL) to advance early detection diagnosis and prognosis decoding in oral cancer and other cancers as well. Here, we perform an extensive review of the related literature on OSCC detection and using deep learning techniques in medical images. The objective is to provide an in-depth overview of the literature on how deep learning has advanced diagnosis modalities for OSCC and place it within the broader landscape involving conventional methods.

Redie et al.~\cite{redie2023oral} investigated Oral cancer detection using a transfer learning-based framework from histopathology images. The dataset contained 5,192 images in total after augmentation with two classes. The VGG19 model performed the highest accuracy of 96.26\%.

Alanazi et al.~\cite{alanazi2022intelligent} discovered intelligent DL enabled oral squamous cell carcinoma detection and classification using biomedical images. The dataset was a public dataset from Kaggle containing 131 images with two classes. A novel IDL-OSCDC model was discovered in this study, and it gave an accuracy of 95\%.

Das et al.~\cite{das2020automated} investigated the automated classification of cells into multiple classes in epithelial tissue of oral squamous cell carcinoma using transfer learning and CNN. The dataset contained 156 slide images of two classes. The traditional CNN model outperformed the transfer learning models with 97.5\% accuracy.

Warin et al.~\cite{warin2022ai} studied AI-based analysis of oral lesions using CNN for early
detection of oral cancer. The dataset contains of 980 oral photographic images with three classes having 365, 315, and 300 images, respectively. The DenseNet-169 achieved the highest f1 score of 0.98.

Panigrahi et al.~\cite{panigrahi2022capsule} investigated capsule network-based analysis of histopathological images of oral squamous cell carcinoma. Their dataset had 150 number of samples. The best model of this work was CapsNet with 97.35\% accuracy.

Mira et al.~\cite{mira2024early} investigated early diagnosis of oral cancer using image processing. The dataset contains a total of 455 samples. A novel DL model was introduced here, and the model got an accuracy of 84.3\%.

Su et al.~\cite{su2021current} studied current insights into oral cancer diagnostics. The dataset contains a total of 455 samples. Their best model was HRNet-W18, which achieved 84.3\% accuracy.

Welikala et al.~\cite{welikala2020automated} investigated automated detection and classification of oral lesions using DL. The dataset contains a total of 2155 images. Their best model was ResNet-101, with an 87.03\% f1 score.

Jubair et al.~\cite{jubair2022novel} studied a lightweight CNN for early detection of oral cancer. There were a total of 716 images containing two classes. EfficientNet-B0 was the best model with 85\% accuracy.

Ahmad et al.~\cite{ahmad2023multi} investigated multi-method analysis of histopathological images for early diagnosis of oral squamous cell carcinoma using DL and hybrid techniques. The dataset was the one that we used in this study. It had 5192 images after augmentation, among which 48\% was normal and 52\% infected. Their hybrid model SVM with feature extraction of DesnseNet201 achieved the highest accuracy of 97\%.

Palaskar et al.~\cite{palaskar2020transfer} studied transfer learning for oral cancer detection. The dataset had a total of 1224 images divided into two sets. The best model of this study was InceptionV3, with 91.13\% accuracy and 0.88 f1 score.

\section{Methodology}
\subsection{Dataset}
This project got data mostly from the public dataset \enquote{Histopathological imaging database for oral cancer analysis}~\cite{rahman2020histopathological}. The dataset was accessed from kaggle~\cite{kebede2021dataset}, where images are divided into training, testing, and validation sets accompanied by two labels: Normal and OSCC (Oral Squamous Cell Carcinoma). The total number of images used in the training data is 2435 Normal and 2511 OSCC. The testing data contains 31 Normal images and 95 OSCC. The dataset consists of 5192 images in total, with around 52\% being OSCC and the remaining Normal (about 48\%). The data was split into a 70:30 ratio for training and testing, respectively. 


\subsection{Data Preprocessing}
The resultant images display diversification in their size, level of zoom, angle, lighting condition, and orientation. All images were preprocessed to standardize the dataset for model education. Specifically, they were cropped manually in 224 × 224 and converted to JPG. This preprocessing ensures consistency in the input data format across all images, facilitating compatibility with the chosen deep-learning model architecture. After the preprocessing of the images, they are used for the image augmentation process, such as zooming, right shifting, flipping, etc. 


\subsection{Training Methodology}
This subsection explains the models used in this study in detail.
\subsubsection{ResNet101}
ResNet101 is a convolutional neural network architecture, and it is famous for its depth (how many layers), which makes it really good at image recognition. At 101 layers deep, it outranked the previous winner by using skip connections to solve the vanishing gradients problem. This enables a smooth flow of gradients known as \enquote{unfettered gradient propagation,} which in turn facilitates training very deep networks with eventual optimal convergence and classification accuracy. Residual units are the individual convolutional layers applied within each ResNet101 block. As a further efficiency measure, it uses the bottleneck design of deeper layers without sacrificing performance. ResNet101 models are a common pre-trained model for image classification or other computer vision-related tasks. ResNet101 really pushed deep learning in computer vision through its depth, skip connections and some creative design.
\subsubsection{VGG-16}
VGG-16 is an example of a convolutional neural network that was designed to suit image classification by the Visual Geometry Group (VGG) at the University of Oxford. VGG-16 has 13 convolutional layers and three fully connected layers, a total of 16 weight layers. It accepts an RGB image of 224x224 size and is processed first through convolutional layers followed by a max pooling layer to learn details in images at lower dimensions. Each block contains a few 3x3 convolution layers with ReLU activation to add non-linearity and also increases the number of filters for higher-order patterns. Dropout is used in fully connected layers with 4096 units and ReLU activation to prevent overfitting. The final layer generates the class probabilities using softmax activation. VGG-16 is a very effective model in image classification and has become well-used for research in computer vision.
\subsubsection{DenseNet121}
DenseNet121 is a 121-layer CNN (Densely connected Neural Network) in the family of DenseNets. Eyenet has dense connections between two layers; each unit in a layer is connected to all other units of its succeeding layer (all-to-all connectivity). This makes the network feed-forward fully connected. These are re-used via the propagations and update of gradients, leading to better performance in terms of convergence. The DenseBlock from the dense blocks, which are in each of these dense blocks, has multiple convolutional layers, batch normalization, and ReLu activation functions. The spatial dimension of input and parameters is reduced by dense blocks through transition layers. DenseNet121 mitigates the vanishing gradient problem very well and performs well in tasks such as image classification, object detection, and so on. It has been widely used in research and practical applications thanks to its small size, deployment efficiency (it is pre-trained on ImageNet), and the fact that it makes good use of parameters.
\subsubsection{EfficientNetB3}
EfficientNet family member EfficientNetB3 is one such family of convolutional neural networks that have been widely publicized for their efficiency and performance in image recognition. It was created by Google AI researchers to offer a balanced solution between model size, computational resources, and performance for resource-constrained environments. EfficientNetB3 goes through compound scaling of its depth, width, and resolution to perform better. It consists of several blocks with inverted bottleneck structures, which take advantage of depth-wise separable convolutions and squeeze-and-excitation modules to improve the representational capacity effectively. EfficientNetB3, pre-trained on datasets like ImageNet, is a brilliant transformer for image classification and object detection. Since the design is quite compact and there are fewer parameters, it is a perfect fit for research as well as practical applications in deep learning/computer vision.
\subsection{LIME}
Any black-box classifier may be interpreted by LIME (Local Interpretable Model-Agnostic Explanations), as it is model-agnostic. One of the most popular and straightforward XAI methods is LIME. The statement \enquote{LIME is explainable for any supervised deep learning models in this world} relates to this model agnosticism. A test sample and prediction model are fed into the LIME. It starts with sampling in order to obtain a surrogate dataset. It generates 5000 pseudo-samples and normalizes the project feature vector by default. Through this method, the predictive model is able to attain the target variable for these 5000 samples. Additionally, each row in the surrogate dataset is weighted according to how much the newly produced samples match the original sample~\cite{mahmud2024advance}.
\subsection{Testing Methodology}
This subsection explains how the models were tested for this study. The equations of the testing metrics are given below~\cite{mahmud2024optimizing}:
\begin{equation}
\label{eq:precision}
\text{Precision} = \frac{TP}{TP + FP}
\end{equation}

\begin{equation}
\label{eq:recall}
\text{Recall} = \frac{TP}{TP + FN}
\end{equation}

\begin{equation}
\label{eq:F1}
F1-score= 2 \times \frac{\text{Precision} \times \text{Recall}}{\text{Precision} + \text{Recall}}
\end{equation}

\begin{equation}
\label{eq:accuracy}
\text{Accuracy} = \frac{TP + TN}{TP + TN + FP + FN}
\end{equation}

\subsection{Work Flow Diagram}
\begin{figure}[htbp]
    \centering
    \includegraphics[width=0.4\textwidth]{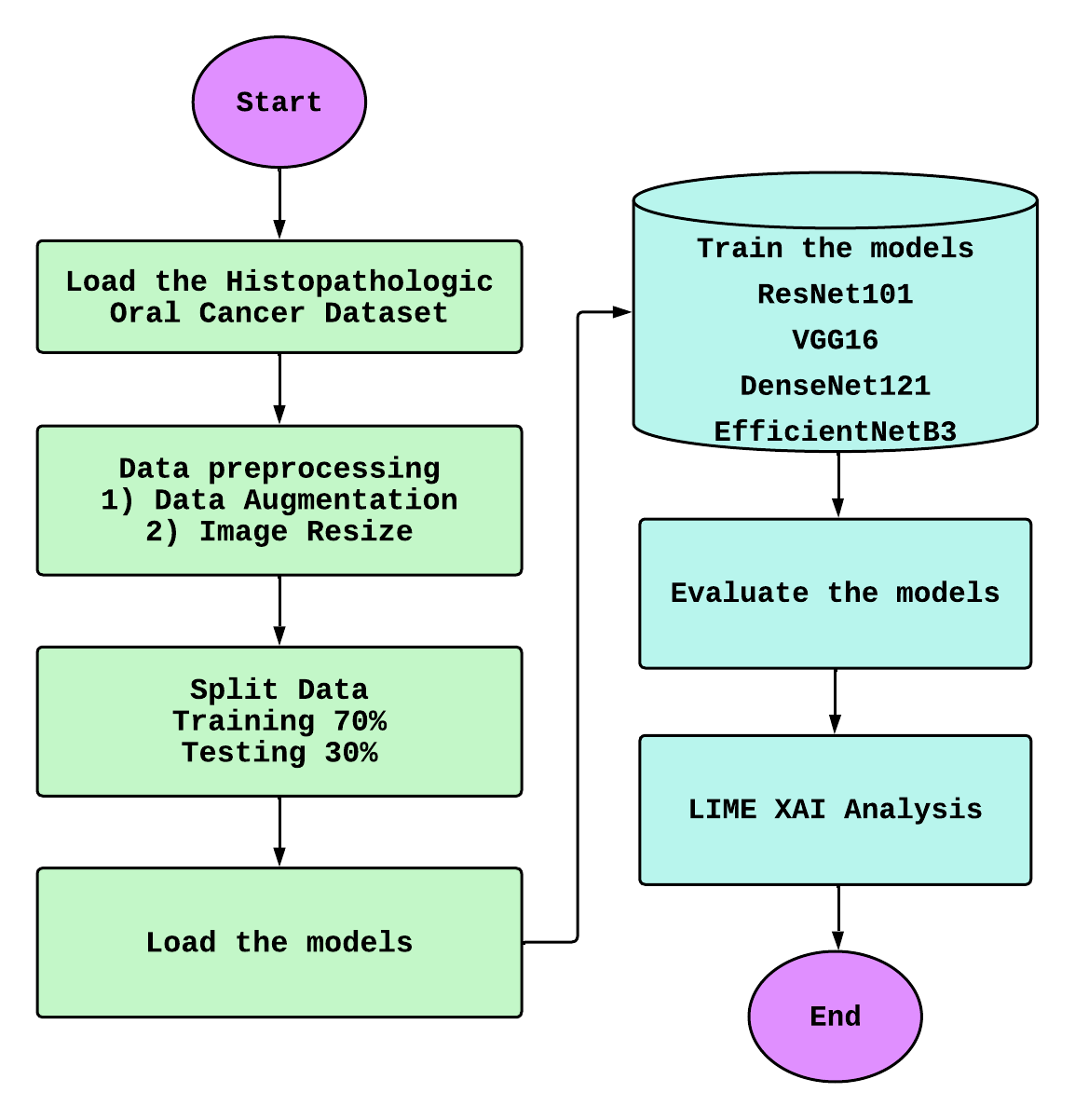}
    \caption{Workflow diagram of this study}
    \label{fig:WF}
\end{figure}

Fig.~\ref{fig:WF} represents the workflow diagram of this study.

\section{Result Analysis}
To classify oral cancer images in our study, we used four different CNN models. Those models are: ResNet101, VGG16, DenseNet121 and EfficientNetB3. The models evaluated various performance metrics, i.e., accuracy, precision, recall, f1-score, confusion matrix, etc. The performance of the best two models is mostly displayed in the upcoming subsections.

\subsection{EfficientNetB3}
\begin{figure}[htbp]
    \centering
    \includegraphics[width=0.4\textwidth]{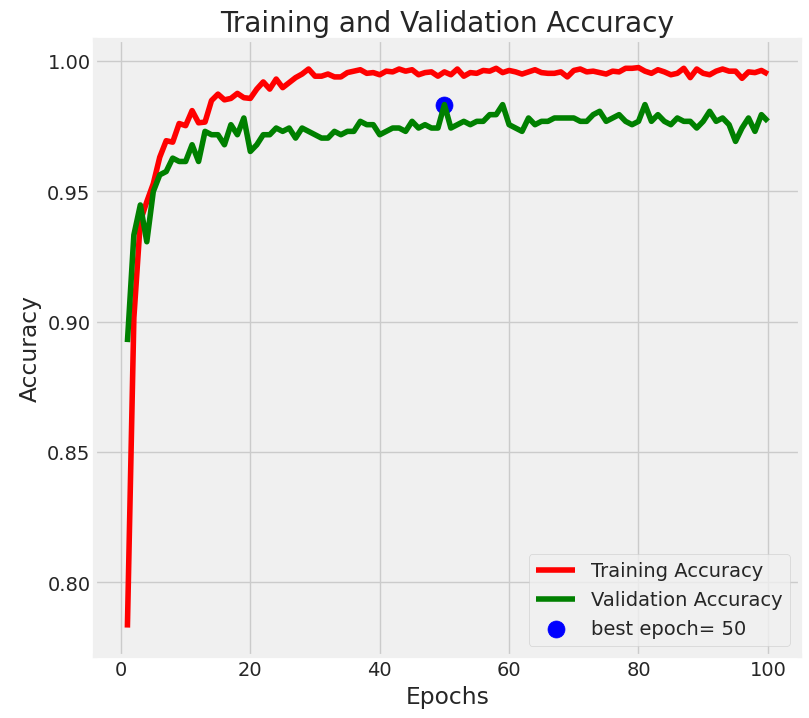}
    \caption{Training and validation accuracy curve of EfficientNetB3}
    \label{fig:EAC}
\end{figure}

Fig.~\ref{fig:EAC} displays accuracy curves of the EfficentNetB3 model for OSCC classification on training and validation datasets. The training accuracy (red) increases quickly and plateaus close to 100\%, suggesting that the model is successfully learning from the training data. The validation accuracy (green) also increases but begins to plateau around 98\%, demonstrating that it generalizes well on unseen data while only slightly diverging from training, which may suggest the early onset of overfitting after epoch 50 - where we previously stated is our \enquote{best} point.

The EfficientNetB3 models of this study all reached the inflection point even before the 20 epoch number for both the training and validation loss curve, as shown in Fig~\ref{fig:ELC}. As we can see below, the training loss (red) and validation loss (green) drop off steeply in the early part of 100 epochs before leveling out — indicating that our model seems to have learned how to minimize error on both train and valid data. The point marked with the best epoch at 74 is where one should stop, beyond which the model may starts over-fitting.

\begin{figure}[htbp]
    \centering
    \includegraphics[width=0.4\textwidth]{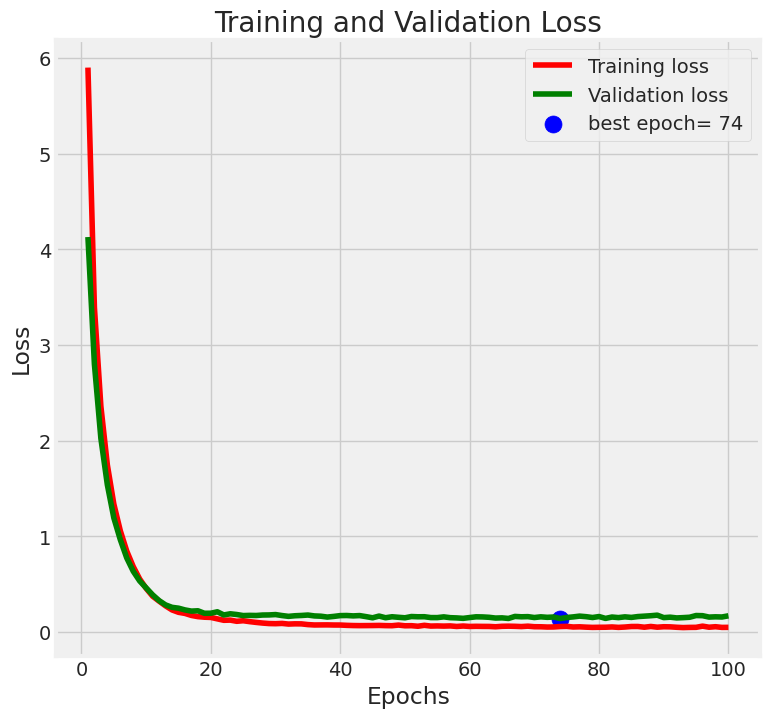}
    \caption{Training and validation loss curve of EfficientNetB3}
    \label{fig:ELC}
\end{figure}

\begin{figure}[htbp]
    \centering
    \includegraphics[width=0.4\textwidth]{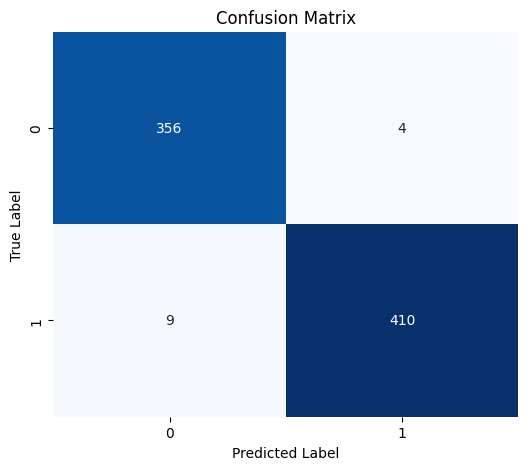}
    \caption{Confusion Matrix of EfficientNetB3}
    \label{fig:CME}
\end{figure}

Fig.~\ref{fig:CME} visualizes the confusion matrix of the EfficientNetB3 model. It is seen that the model classified 356 images correctly as normal, labeled as 0. 410 images were classified correctly as infected, labeled as 1. Only 13 images were classified wrongly. It shows the model's robustness in classifying the disease.

\subsection{DenseNet121}
\begin{figure}[htbp]
    \centering
    \includegraphics[width=0.4\textwidth]{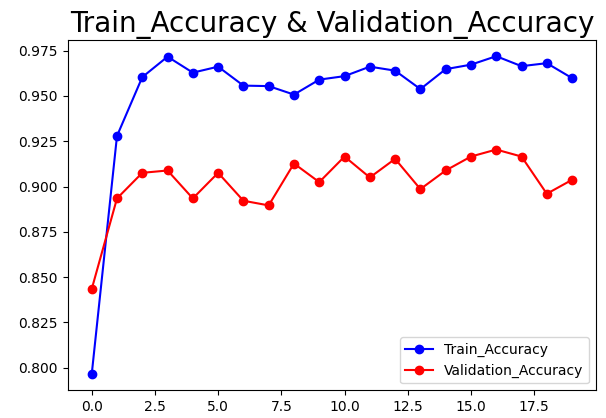}
    \caption{Training and validation accuracy curve of DenseNet121}
    \label{fig:DAC}
\end{figure}

Fig.~\ref{fig:DAC} indicates the training and validation accuracy curves reported for the DenseNet121 model. The blue line for the training accuracy initially starts out low but quickly rises within just a few epochs to stabilize at around 96\% and remains fairly consistent over all of the remaining epochs. However, as visible by the validation accuracy in red, it climbs sharply well but fluctuates and trends lower after peaking around epoch 5.

\begin{figure}[htbp]
    \centering
    \includegraphics[width=0.4\textwidth]{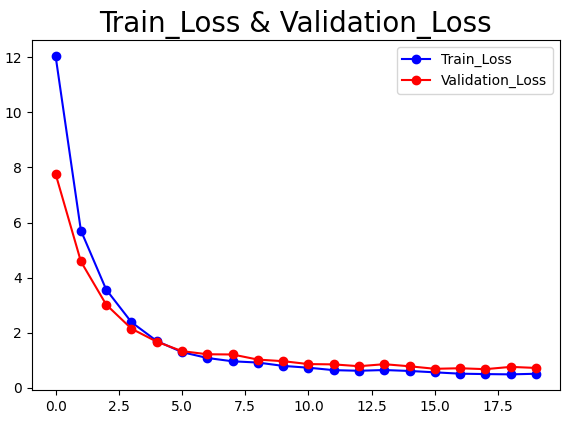}
    \caption{Train and validation loss curve of DenseNet121}
    \label{fig:DLC}
\end{figure}

In Fig.~\ref{fig:DLC}, results of training and testing loss curves are shown for the DenseNet121 model. The training loss (blue) and validation loss (red) both decrease rapidly at first, which means that the model learns very quickly during the early stages of training. Following the first few epochs, both loss curves eventually converge and become very similar to each other — they keep these values stable and low in later runs.

\begin{figure}[htbp]
    \centering
    \includegraphics[width=0.4\textwidth]{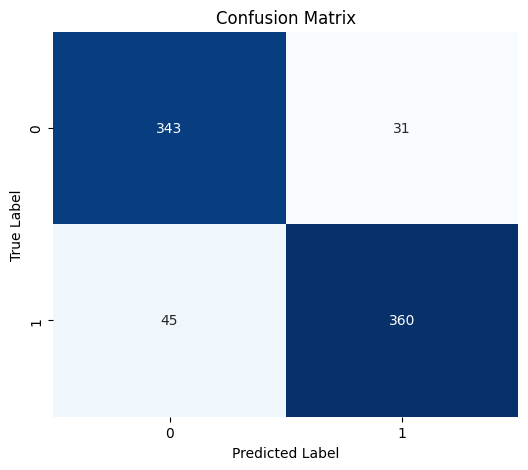}
    \caption{Confusion Matrix of DenseNet121}
    \label{fig:CMD}
\end{figure}

The confusion matrix of the DenseNet121 model was displayed in Fig.~\ref{fig:CMD}. This model didn't perform as well as the earlier one. It classified 343 images correctly as normal, labeled as 0. 360 images were classified correctly as infected, labeled as 1. A total of 76 images were classified wrongly. It shows the model didn't perform as well as the earlier one for classifying this disease.

\subsection{Model Evaluation}
Table~\ref{tab:RE} evaluates the applied models in terms of training accuracy, validation accuracy, and validation loss. It is apparent from the table that the EfficientNetB3 model worked better than all other applied models in this dataset, with 97.59\% validation accuracy and validation loss of only 0.1117. The DenseNet121 also performed well with 91\% validation accuracy and 0.3511 validation loss.

\begin{table}[htbp]
 \scriptsize
\caption{Result Evaluation of Models}
\centering
\begin{tabular}{|c|c|c|c|}
\hline
\textbf{Models} & \textbf{Training Accuracy} & \textbf{Validation Accuracy} & \textbf{Validation Loss} \\
\hline
EfficientNetB3  & 0.9964 & 0.9759 & 0.1117\\
\hline
DenseNet121  & 0.9600 & 0.9100 & 0.3511\\
\hline
ResNet101  & 0.9343 & 0.9058 & 0.2535 \\
\hline
VGG16 & 0.8647 &  0.8314 & 0.4515 \\
\hline
\end{tabular}
\label{tab:RE}
\end{table}

Table~\ref{tab:PE} represents the performance metrics of the models and also compares those. It is evident here also that EfficientNetB3 outperformed all other models with an excellent accuracy of 98.33\%. The model had a precision of 0.9903, a recall of 0.9782, and an f1 score of 0.9844. It only used 12M parameters to achieve these metrics. DenseNet121, with fewer parameters, i.e., 8M, came out second by achieving an accuracy of 90.45\%.

\begin{table}[htbp]
 \scriptsize
\caption{Performance Metrics Comparison}
\centering
\begin{tabular}{|c|c|c|c|c|c|}
\hline
\textbf{Models} & \textbf{Parameters} & \textbf{Precision} & \textbf{Recall} & \textbf{F1-Score} & \textbf{Accuracy}\\
\hline
EfficientNetB3 & 12M&0.9903 &0.9782 &0.9844 &0.9833\\
\hline
DenseNet121  & 8M&0.9207 &0.8889 &0.9045 &0.9024\\
\hline
ResNet101 &44.5M  &0.9410 &0.8947 &0.9173 &0.8931\\
\hline
VGG1616 &138M& 0.9036 &0.7895 &0.8427 &0.8177\\
\hline
\end{tabular}
\label{tab:PE}
\end{table}

\subsection{Result Comparison}
Table~\ref{tab:RC} shows the comparison of the applied models with the existing ones by others. It is observed that EfficientNetB3 outplayed all existing models with an accuracy of 98.33\%.

\begin{table}[htbp]
\caption{Result Comparison}
\centering
\begin{tabular}{|c|c|c|c|c|c|}
\hline
\textbf{Study}  & \textbf{Model} & \textbf{Accuracy (\%)} & \textbf{F1 score (\%)} \\
\hline
This study   & EfficientNetB3 & \textbf{98.33} & \textbf{98.44}\\
\hline
\cite{warin2022ai} & DenseNet-169 & N/A & 98.00 \\
\hline
\cite{das2020automated} & CNN  & 97.50 & N/A\\
\hline
\cite{panigrahi2022capsule} & CapsNet &97.35& N/A \\
\hline
\cite{redie2023oral}  &VGG19 & 96.26 & N/A\\
\hline
\end{tabular}
\label{tab:RC}
\end{table}

\noindent \textbf{N.B: N/A represents Not Available}

\subsection{Explainable AI}
\begin{figure}[htbp]
    \centering
    \includegraphics[width=0.4\textwidth]{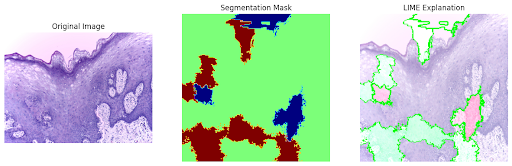}
    \caption{LIME XAI visualization for the EfficientNetB3 model on an infected image}
    \label{fig:XAI_1}
\end{figure}

\begin{figure}[htbp]
    \centering
    \includegraphics[width=0.4\textwidth]{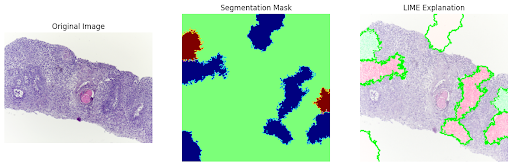}
    \caption{LIME XAI visualization for the EfficientNetB3 model on a non-infected image}
    \label{fig:XAI_2}
\end{figure}

In this study, EfficientNetB3 came out to be the best model among the applied ones. For that reason, we decided to apply LIME XAI in this model. We have randomly selected two images, among which the most image of Fig.~\ref{fig:XAI_1} is an image with OSCC, and the most image of Fig.~\ref{fig:XAI_2} is an image without OSCC. The segmentation masks for the various areas that the model learns from each supplied picture are displayed in the center row. More crucial sections are shown by the black area in the mask, indicating that this leads to a low rating. Higher-intensity areas in the mask denote areas that have a bigger impact on the model's prediction. The figures illustrate how the rightmost images on both sides draw attention to the information and places that our model was lucky enough to attract. This suggests that in order to differentiate between photos with and without OSCC infection, texture, and color in these areas are crucial.

\section{Conclusion and Future Work}
This study demonstrates the considerable promise of deep learning methods in histopathological image analysis and holds importance for OSCC diagnosis. Although all tested models performed well, EfficientNetB3 stood out with a 98.33\% accuracy score and an F1 score of 98.44\%. It is a computationally efficient model, which makes it an appealing one for real-time medical diagnostics. DenseNet121 did get the right answer a step-down, which also indicates that deep-learning models are performing well with their decision-making and can handle complications in medical domain image data. Secondly, we also incorporated Local Interpretable Model-agnostic Explanations (LIME) along with EfficientNetB3 to explain the model. One of the important points that touches on explainable AI (XAI) techniques, which are indispensable for boosting trust and reliability in an application intended to simulate medical diagnostics based solely on image data is the ability of human experts — doctors with experience at making predictions from radiological images to understand what makes a model predict. These findings imply the potential of deep learning integrated with XAI in improving OSCC diagnostic accuracy and efficiency, which could enable earlier, more robust detection. The next steps should involve improving these models in terms of performance and application to other cancer types, along with images from Bangladeshi hospitals. Also, different explainable AI techniques will be used to understand the model decision-making more precisely.

\bibliographystyle{IEEEtran}
\bibliography{IEEE.bib}

\end{document}